# Study of shape coexistence in the $^{180-190}$Hg isotopes by *SO*(6) representation of eigenstates

H. Sabri*, A. Ghale Asadi, O. Jabbarzade, S. K. Mousavi Mobarake

Department of Physics, University of Tabriz, Tabriz 51664, Iran.

* E-mail address: h-sabri@tabrizu.ac.ir

**Abstract**

In this paper, we have studied the shapes coexistence in the $^{180-190}$Hg isotopes. The *SO*(6) representation of eigenstates and a transitional Hamiltonian in the Interacting Boson Model are used to consider the evolution from prolate to oblate shapes for systems with total boson number $N = 9 - 12$. Parameter free (up to overall scale factors) predictions for energy spectra and quadrupole transition rates are found to be in good agreement with experimental counterparts. The results for the control parameter of transitional Hamiltonian offer a combination of spherical and deformed shapes in these Hg isotopes and also more deviation from SO(6) limit is observed when the quadrupole deformation is decreased. Also, there are some suggestion about the expectation values of $\hat{n}_d$ operator which are determined in the first state of ground, beta and gamma bounds and the control parameter of model.

## 1. Introduction

The microscopic origin of quadrupole collectivity and shape coexistence at low excitation energies in neutron mid-shell nuclei near the $Z = 50$ and 82 shell closures are still not fully understood. In some of these nuclei, the deformed intruder states coexist with the nearly spherical normal states [1-25]. The neutron mid-shell $Z \approx 82$ nuclei are very neutron deficient lying close to the proton drip line and they can be produced in fusion-evaporation reactions, albeit, due to strong fission competition. Very important information about the shape-coexisting states in this region has been extracted in *α*-decay studies of fusion products, especially when detecting *γ* rays or electrons in coincidence with *α* particles [20-47].

There are two naturally complementary ways in order to describe the phenomenon of nuclear shape coexistence [45-62]. The first one is model that starts from a nuclear shell-model approach, protons and neutrons are expected to gradually fill the various shells at $Z, N = 2, 8, 20, 28, ...$ giving rise to a number of double-closed shell nuclei that are the reference points determining shells. The valence nucleons in these shells have been allowed to interact through either a phenomenologically fitted effective interaction or a microscopic effective interaction, deduced from many-body theory from realistic NN forces. In the other approach, the starting point is an effective nuclear force or energy-density functional which are used to derive the optimized single-particle basis in a self-consistent way. These approaches use of Hartree-Fock (HF), or using Hartree-Fock-Bogoliubov (HFB) theory, when also including the strong nucleon pairing forces in both cases constraining the nuclear density distribution to specific values for the quadrupole moments, octupole moments, etc. [31-40].

In the Interacting Boson Model (IBM) [10-35], which describes the nuclear structure of even–even nuclei within the $U(6)$ symmetry, possessing $U(5)$, $SU(3)$ and $O(6)$ dynamical symmetry limits, shape phase transitions were studied 25 years ago with using the classical limit of the model. These descriptions point out that there is a first order shape phase transition between $U(5)$ and $SU(3)$ limits, namely between spherical and deformed limits which $Hg$ isotopes are expected to lie in this transitional region. The analytic description of nuclear structure at the critical point of phase transitions has attracted extensive interest in the recent decades. One has to employ some complicated numerical methods to diagonalize the transitional Hamiltonian in these situations but we have proposed a new solution which was based on the $SO(6)$ representation of eigenstates in this transitional region and has applied to some nuclei with total boson number $N = 3 - 5$ [21-22].

In this paper, we have considered the shape coexistence phenomena in the $^{180-190}$Hg isotopes. A simple Hamiltonian with two control parameters was used. We have determined the *SO*(6) representation [22-30] of eigenstates and by using them, the matrix elements of quadrupole term in Hamiltonian are determined for systems with total boson number $N$ = 9-12. The parameters of the Hamiltonians are fixed through a least square fit to the known energies and absolute B(E2) transition rates of states up to 3 MeV. Parameter free prediction for energy spectra and transition rates have compared with the most recent available experimental data [43-48] for these nuclei which a fairly good agreement is achieved.

2. The model

2.1. investigation of shape coexistence by other models

As have been shown in various spectroscopic selective experiments, *e.g.* transfer reactions in particular, very near to closed shells (the In and Sb nuclei at Z=50 but also in other mass regions, *e.g.* the Tl and Bi nuclei at Z=82) some low-lying extra states, so-called intruder states, have been observed with a conspicuous energy dependence on the number of free valence neutrons, hinting for 2*p*-2*h* excitations as their origin [49-61]. If these excitations are proton excitations combined with the neutron degree of freedom appearing on both sides of the Z=50 closed shell, such as condition which are available for Te isotopes, it is a natural step to suggest that low-lying extra 0$^+$ excitations will also show up in the even-even nuclei in between. Because the Te isotopes with a large number of valence neutrons are situated near to the $\gamma$-stability line, they could be studied [62-65].

Since making a 1*p*-1*h* excitation across the closed shell at Z=50 takes about 4.5 MeV (the proton shell gap), the unperturbed energy for 2p-2h excitations comes up to about 9 MeV. Even though pairing amongst the particles and holes will lower the energy in an important way to 4-5 MeV, this is still far from the observed excitation energy of 1.7 MeV. Some essential element is missing when starting from the spherical intrinsic symmetry of the shell model.

One way to come around is breaking the spherical symmetry and allowing the mean field to acquire quadrupole deformation thereby giving rise to the possibility that spherical orbits split and the large spherical shell gap at Z=50 and also at Z=82 rapidly vanishes. Calculations have been carried out over the years using deformed mean-field studies, *e.g.* Nilsson model, deformed Woods-Saxon, Hartree-Fock-Bogoliubov studies and we would like to cite in particular [65-72]. The deformed field essentially points out to need for the quadrupole component in the mean field as the agent for the increased binding energy. Knowing this, and having experimental knowledge of the fact that 1*p*-1*h* (in odd-mass nuclei) and 2*p*-2*h* (in even-even nuclei) are present in these states, it is tempting to incorporate this in a spherical shell-model description. By invoking a schematic model that was discussed in detail in [72-75] it is possible to evaluate the excitation energy of a 2*p*-2*h* configuration:

$$E_{intr.}(0^+) = \langle 0_I^+|H|0_I^+\rangle - \langle 0_{GS}^+|H|0_{GS}^+\rangle \quad ,$$

in which the index *I* denotes the nucleon distribution in the intruder state and *GS* the distribution in the ground state. Using a pair distribution for the neutrons, combined with a 2*p*-2*h* excitation and a 0*p*-0*h* excitation for the intruder and regular state, respectively, one can derive the expression:

$$E_{intr.}(2p-2h) \cong 2(\varepsilon_p - \varepsilon_h) + \Delta E_M(2p-2h) - \Delta E_{pair} + \Delta E_Q(2p-2h) \quad ,$$

where the various terms describe the unperturbed energy to create the 2*p*-2*h* configuration. A monopole correction due to a change in proton single-particle energy while changing the neutron number, the pairing-energy correction because $0^+$- coupled pairs are formed, and the quadrupole binding energy originating from the proton-neutron force, respectively.

In calculating the neutron number dependence of the 2p-2h intruder $0^+$ configurations, we have to determine the quadrupole energy contribution and this we do by using the SU(3) expression given in [72-75], *i.e.*:

$$\Delta E_Q \cong 2\kappa \Delta N_\pi N_\nu,$$

in which $\Delta N_\pi$ denotes the number of pairs excited out of the closed shell configuration at Z=50, *i.e.*, $\Delta N_\pi = 2$ for a 2p-2h excitation.

This approach points out that the essential elements are the strong pairing interactions amongst the particles and the holes that make up for the excited configuration and the strong quadrupole proton-neutron forces. It is precisely here that early contacts between the disconnected "spaces" of interacting boson within a valence space only and the *p-h* excitations of the core itself showed up. In a lowest order approximation, one can think of the 2*p* and the 2*h* parts to bring in two extra bosons increasing the active model space from N to *N+2* bosons and carry out separate calculations for both spaces introducing a coupling between them by using a mixing Hamiltonian

$$H_{mix} = \alpha \left(s^\dagger s^\dagger\right)^0 + \beta \left(d^\dagger d^\dagger\right)^0 + h.c.$$

The presence of these extra states, characterized by 2*p*-2*h* excitations across the Z=50 shell closure, has become a fingerprint especially near the N=66 mid-shell region. Moreover, the interference between the regular vibrational states and these intruder states that contain a much larger collectivity, shows up as drastic modifications of the regular vibrational E2 intensity ratios.

A full shell-model study of the Te nuclei, with neutrons moving all through the full valence space of N=50 towards N=82, at the same time incorporating, besides the two proton holes outside of Z=50, the 2*p*-2*h* excitations that show up in the mid-shell neutron region (around N~66), is out of reach. Only when approaching the neutron shell closure at N=50 i.e. N=52 and 54 and for the heavy nuclei near N=82 considering the cases with N=78-80 and also beyond, at N=84, full shell-model studies can become feasible. Therefore, the study of these extreme heavy nuclei is important since it may shed light on the way how collective quadrupole states (with an harmonicities included) may go over into the shell-model structure: there should be some region of overlap which can give us very interesting information.

As mentioned in Refs.[72-88], the nuclear shell model is not in a position to be used for a reliable computation of the low-energy properties of the full range of Hg isotopes. This means that one has to resort to a suitable truncation of the shell model, such a model has been developed in Ref.[74] which is based on quasiparticle random-phase approximation (QRPA) or use algebraic approach to explore the considered nuclei. On the other hand, the drawback of those calculations is that one easily gets involved with a lot of parameters and unless one has some physics guidance the detailed agreement needs some caution. In the following, we have developed our previous algorithm [22] to get all SO(6) representation of a two parameter Hamiltonian which make a very simple method to consider shape evolution and coexistence[70-85].

2.2. Transitional Hamiltonian and $SO(6)$ representation

Phase transitions have been studied widely in Refs.[ 10-19, 22-30] are those of the ground state deformation. In the Interacting Boson Model (IBM), one would achieve a very simple two parameters description leading to a symmetry triangle which is known as extended Casten triangle. There are four dynamical symmetries of IBM called U(5), SU(3), $\overline{SU(3)}$ and O(6) limits. They correspond to vibrational nuclei with a spherical form, i.e. U(5), an axially symmetric prolate rotor with a minimum in the energy at $\gamma = 0°$ which corresponds to SU(3) and an axially symmetric oblate rotor with a minimum at $\gamma = 60°$, namely $\overline{SU(3)}$. The fourth symmetry is located in the middle of the $SU(3) \leftrightarrow \overline{SU(3)}$ transitional region and corresponds to a rotor with a flat potential in $\gamma$, e.g. O(6) limit [23-25]. On the other hand, in the Bohr-Mottelson Collective model framework, Bonatsos *et al*, has introduced the Z(5) critical point symmetry for the prolate to oblate shape phase transition. It is known many predictions of this model which involving large rigid triaxiality, are very close to the predictions of $\gamma$ – soft models involving $\gamma$ – fluctuation such that, $\gamma_{rigid}$ of the former equals of the latter. Also, equivalence between $\gamma$ – instability and rigid triaxiality with $\gamma = 30^{\circ}$ has been shown in relation to O(6) limit of IBM. To consider this transitional region, it is parameterized using simple Hamiltonian as has been introduced in Refs.[12-14,22]:

$$\hat{H}(N,\eta,\chi) = E_0 + \eta \hat{n}_d + \frac{\eta - 1}{N} \hat{Q}_\chi . \hat{Q}_\chi + CL^2 \quad , \quad (1)$$

Where $\hat{n}_d = d^\dagger . \tilde{d}$ is the $d$ – boson number operator and $\hat{Q}_\chi = (s^\dagger \tilde{d} + d^\dagger s)^{(2)} + \chi(d^\dagger \times \tilde{d})^{(2)}$ represents the quadrupole operator and $N(= n_s + n_d)$ stands for the total number of bosons. Also, the $\eta$ and $\chi$ quantities are regard as control parameters and can vary within the range $\eta \in [0,1]$ and $\chi \in [-\sqrt{7/2}, +\sqrt{7/2}]$. The transitional region in this study, namely the prolate-oblate transitional region, passing through the O(6) dynamical symmetry limit, is known to be situated close to the upper right leg of the extended Casten triangle with $\eta = 0$.

In the following, we have employed SO(6) representation to determine the eigenvalues of Hamiltonian (2). Algebraic structure of IBM has been described in detail in Refs.[22-26] and especially in Ref.[28]. Here, we briefly outline the basic ansatz and summarize the results which have been used in this paper for our considered representation. Classification of states in the *SO*(6) representation is [27-29]:

$$U(6) \supset SO(6) \supset SO(5) \supset SO(3) \supset SO(2) \quad , \qquad (2)$$
$$\downarrow \quad\quad \downarrow \quad\quad \downarrow \quad\quad \downarrow \quad\quad \downarrow$$
$$[N] \quad\quad \langle\Sigma\rangle \quad\quad (\tau) \quad\quad L \quad\quad M$$

The multiplicity label $v_\Delta$ in the $SO(5) \supset SO(3)$ reduction will be omitted in the following when it is not needed. Eigenstates $|[N]\langle\sigma\rangle(\tau)v_\Delta LM\rangle$ are obtained with a Hamiltonian with SO(6) dynamical symmetry. Construction of our considered representation requires $n$-boson creation and annihilation operators with definite tensor character in the basis (2) as;

$$B^\dagger_{[n]\langle\sigma\rangle(\tau)lm} \quad , \quad \tilde{B}_{[n]\langle\sigma\rangle(\tau)lm} \equiv (-1)^{l-m}(B^\dagger_{[n]\langle\sigma\rangle(\tau)l,-m})^\dagger \quad , \qquad (3)$$

Of particular interest are tensor operators with $\sigma < n$. They have the property:

$$\tilde{B}_{[n^5]\langle\sigma\rangle(\tau)lm}\big|[N]\langle N\rangle(\tau)v_\Delta LM\big\rangle = 0 \quad , \qquad \sigma < n \qquad (4)$$

For all possible values of $\tau$ and $L$ which are contained in the SO(6) irrep $\langle N\rangle$. This is so because the action of $\tilde{B}_{[n^5]\langle\sigma\rangle(\tau)lm}$ leads to an $(N-n)$-boson state which contains the SO(6) irrep $\langle\Sigma\rangle = \langle N-n-2i\rangle$, $i = 0,1,...$, which cannot be coupled with $\langle\sigma\rangle$ to yield $\langle\Sigma\rangle = \langle N\rangle$, since $\sigma < n$. Number conserving normal ordered interactions that are constructed out of such tensors with $\sigma < n$ (and their Hermitian conjugates), thus have $|[N]\langle N\rangle(\tau)v_\Delta LM\rangle$ as eigenstates with zero eigenvalues. A systematic enumeration of all interactions with this property is a simple matter of SO(6) coupling. For one body operators,

$$B^\dagger_{[1]\langle 1\rangle(0)00} = s^\dagger \equiv b^\dagger_0 \quad , \qquad B^\dagger_{[1]\langle 1\rangle(1)2m} = d^\dagger_m \equiv b^\dagger_{2m} \quad , \qquad (5)$$

On the other hand, coupled two body operators are of the form:

$$B^\dagger_{[2]\langle\sigma\rangle(\tau)lm} \propto \sum_{\tau_k \tau_{k'}} \sum_{kk'} C^{\langle\sigma\rangle(\tau)l}_{\langle l\rangle(\tau_k)k,\langle l\rangle(\tau_{k'})k'} (b^\dagger_k b^\dagger_{k'})^{(l)}_m \quad , \qquad (6)$$

where $(b^\dagger_k b^\dagger_{k'})^{(l)}_m$ represent coupling to angular momentum $(l)$ and the $C$ coefficients are known $SO(6) \supset SO(5) \supset SO(3)$ isoscalar factors. These processes lead to the normalized two-boson

SO(6) tensors which are displayed in Tables 1- 4 for systems with total boson number $N = 9$ to 12, respectively. There is one operator with $\sigma < n = 2$ and it gives rise to the following SO(6) - invariant interaction;

$$\hat{B}^{\dagger}_{[2]\langle 0\rangle(0)00} \tilde{B}_{[2^5]\langle 0\rangle(0)00} = \frac{1}{3}\hat{P}_+\hat{P}_-$$

which is simply the SO(6) term in $\hat{H}_{DS}$, Eq. (1). This proves that a two-body interaction which is diagonal in $|[N]\langle N\rangle(\tau)v_\Delta LM\rangle$ is diagonal in all states $|[N]\langle\Sigma\rangle(\tau)v_\Delta LM\rangle$ [13].

Now, with using these eigenstates, the energy spectra for considered systems are determined as:

$$\langle [N]\langle\sigma\rangle(\tau)v_\Delta LM|H|[N]\langle\sigma\rangle(\tau)v_\Delta LM\rangle = E_0 + \eta n_d + \frac{\eta-1}{N}\varepsilon + CL(L+1) \qquad , \qquad (7)$$

The $\varepsilon$ term in Eq.(7) denotes the analytical expressions of quadrupole term in Hamiltonian as presented in Tables 5 - 8 for systems with total boson number $N$ = 9-12 . To get these expressions, we have to diagonalize the matrix of Hamiltonian in these states. To this aim, we have devolved a method which have used in Ref.[22]. In this method, we have blocked the matrix of Hamiltonian in the 3*3 dimensions and then, combine results which finally yield the indicated results for $\varepsilon$. On the other hand, we have used a numerical method, based on MATLAB software, to determine the constants of relation 7, namely $E_0$, $\eta$ and $C$. We have used the recent empirical data for energy spectra and transition probabilities, which have explained in the following, of considered isotopes. The results are presented in the captions of Figures 1a-1f.

2.2. B(E0) and B(E2) Transition Probabilities

The reduced electric monopole and quadrupole transition probabilities are considered as the observables which as well as the quadrupole moment ratios within the low-lying state bands prepare more information about the nuclear structure. The most general one-body multipole transition operator has the form [5,7]:

$$T^{(l)} = q_2\delta_{l2}(d^\dagger \times s + s^\dagger \times d)^{(2)}_\rho + p_l(d^\dagger \times d)^l + r_0\delta_{l0}(s^\dagger \times s)^{(0)}_\rho \quad , \tag{8}$$

where $\rho = \pi$ or $\nu$. Also, in IBM-2 formalism we consider separate terms for proton and neutron, and $\delta_{l2}$ (and $\delta_{l0}$) are Kronecker deltas. Also $q$, $p$ and $r$ are the constants which are extracted from the experimental data and $s^\dagger(d^\dagger)$ represent the creation operator of $s(d)$ boson. The T(E0) operator may be found by setting $l=0$ in above equation as[5,7]:

$$T^{(0)} = p_0(d^\dagger \times d)^{(0)}_\rho + r_0(s^\dagger \times s)^{(0)}_\rho \quad , \tag{9}$$

On the other hand, the E*2* transition operator must be a Hermitian tensor of rank two and consequently, number of bosons must be conserved. With these constraints, there are two operators possible in the lowest order, therefore the electric quadrupole transition operator employed in this study is defined as [7],

$$\hat{T}^{(E2)}_\mu = q_2 \, [\hat{d}^\dagger \times \tilde{s} + \hat{s}^\dagger \times \tilde{d}]^{(2)}_\mu + p_2 \, [\hat{d}^\dagger \times \tilde{d}]^{(2)}_\mu \quad , \tag{10}$$

To evaluate the B(E0) and B(E2) transition ratios and consider the effect of intruder states, we have calculated the matrix elements of T(E0) and T(E2) operators between the considered states which are labeled as our model formalism and then, we can extract the constant quantities of Eqs.(9) and (10) in comparison with empirical evidences. Now, with using SO(6) representation of eigenstates and method has been introduced in Refs.[2-4], the monopole and quadrupole transition rates are determine in the $SU(3) \leftrightarrow \overline{SU(3)}$ transitional region. Similar to energy spectra, significant variations in transition probabilities, propose a structural changes in nuclear structure which can be considered as phase transition between these limits. On the other hand, Jolie *et al.* have predicted in Refs.[31-32], $B(E2; 2^+_2 \to 2^+_1)$ value should has a peak with a collective value which counterpart with O(6) dynamical symmetry and then, decrease quickly as $|\chi|$ increases.

Our results suggest similar behavior for this quantity in considered transition region which more details will present in the following.

On the other hand, we can use the E0 and E2 transition probabilities to consider the evolution of charge radii and spectroscopic quadrupole moments of the $2^+$ states in the ground and first excited bands. As have described in Refs,[89-90], the mean-square charge radius of a state $|s\rangle$ is given by:

$$\langle r^2 \rangle_s = \langle s | \hat{T}_{E0}(r^2) | s \rangle = \frac{1}{e_n N + e_p Z} \left\langle s \left| \sum_{k=1}^{A} e_k r_k^2 \right| s \right\rangle \quad (11)$$

which $e_p$ and $e_n$ denote the effective charges for the neutrons (n) and protons (p). In the IBM-1 the charge radius operator is taken as the most general scalar expression, linear in the generators of U(6) [90],

$$\hat{T}(r^2) = \langle r^2 \rangle_c + \alpha N_b + \eta \frac{\hat{n}_d}{N_b}$$

where $N_b$ is the total boson number, $\hat{n}_d$ is the $d$-boson number operator, and $\alpha$ and $\eta$ are parameters with units of length2. The first term of this equation, $\langle r^2 \rangle_c$, is the square of the charge radius of the core nucleus. The second term accounts for the (locally linear) increase in the charge radius due to the addition of two nucleons (*i.e.*, neutrons since isotope shifts are considered in this study). The third term in Eq. (7) stands for the contribution to the charge radius due to deformation. The factor $1/N_b$ is included here because it is the fraction $\frac{\hat{n}_d}{N_b}$ which is a measure of the quadrupole deformation ($\beta_2$ in the geometric collective model) rather than the matrix element $\hat{n}_d$ itself. Although the coefficients $\alpha$ and $\eta$ will be treated as parameters and fitted to data on charge radii and E0 transitions, it is important to have an estimate of their order of magnitude. The term in *α* increases with particle number and therefore can be associated with the "standard" isotope shift. On the other hand, The term in *η* stands for the contribution to the nuclear radius due to deformation. We have followed the prescription that introduced in Refs.[89-90] by Zerguine et al, and got $\alpha \sim 0.223$ fm$^2$ and *η* values between 0.33 and 0.82 fm$^2$. Also, we have used our results about quadrupole transition probabilities to determine the quadrupole moments of the 2$^+$ states in the ground and first excited bands. The spectroscopic quadrupole moment $Q_s$ is related to the intrinsic quadrupole moment $Q_0$ by the relation [13],

$$Q_s = Q_0 \frac{3K^2 - J(J+1)}{(J+1)(2J+3)}$$

where *K* and *J* stands to describe the band and level which the quadrupole moments are determined. The intrinsic quadrupole moment of nucleus is related to quadrupole transition probability as:

$$B(E2; KJ_1 \to KJ_2) = \frac{5}{16\pi} e^2 Q_0^2 \langle J_1 K 20 | J_2 K \rangle^2.$$

A comparison between the theoretical prediction and experimental counterparts for charge radii and quadrupole moments are presented in Table 9.

3. Results and Discussion

- Theoretical results and comparison with empirical counterparts

We have studied the energy spectra and quadrupole transition rates of $^{180-190}_{80}$Hg isotopes with emphasis on the signatures of shape coexistence. Theses nuclei have been interpreted as the best candidates for coexistence of spherical and deformed shapes which are investigated by different methods such as configuration mixing IBM [11-14], Total-Routhian-surface (TRS) calculations and a symmetry-based approach [14-18].

- Energy spectra

We have determined the low-lying part of energy spectra by employing Eq.(7) and the analytical expression of quadrupole term which are introduced in Tables 5-8 for considered isotopes as have displayed in the Figure 1. Also, we have extracted the constants of Hamiltonian, namely $E_0$, $\eta$, $\chi$ and $C$ by least square fit to the experimental data [43-48] for energy levels and absolute transition probabilities. A general agreement between the theoretical results and experimental counterparts is achieved.

Energy spectra which obtained in this approach are generally in good agreements with the experimental data and indicate the elegance of extraction procedure which presented in this technique and they suggest the success of guess in parameterization. Also, our results for $\eta$ values, the control parameter of transitional Hamiltonian, are compared with the experimental quadrupole deformation values. Results in Figure 2 show an obvious relation between these quantities where nuclei with more deformation, have the biggest $\eta$ values and theoretical predictions suggest an approach to SU(3) limit for them.

Our first assumption for $\eta$ values, which expect to have zero values for these nuclei that are located in or near the critical point of the oblate to the prolate transitional region, is changed and we got nonzero values for this quantity by extraction processes. If we get the $\eta$ values zero in our calculation and determine the variation of our theoretical results in comparison with experimental equivalents as $\sigma = (\frac{1}{N_{tot}} \sum_{i,\,tot} | E_{\exp}(i) - E_{cal}(i) |^2)^{1/2}$, where $N_{tot}$ is the number of energy levels included in the extraction processes, the uncertainty of theoretical predictions are increased obviously.

These results for the $\eta$ in different isotopes describe the effect of spherical shape on the deformed one for these isotopes. As have mentioned in different literatures [12-14, 72-88], the Hg isotopic chain expect to be located in and near the critical point of prolate to oblate transitional region, e.g. SO(6) dynamical symmetry. The results which offer the role of spherical symmetry in this isotopic chain and consequently the combination of these two symmetries, not absolutely but likely, may suggest a shape coexistence-like meaning. Also, if we consider that, interplay between the stabilizing effect of a closed shell on one hand and the residual interactions between protons and neutrons outside closed shells on the other hand, leads to the concept of 'shape coexistence', where normal near-spherical and deformed structures coexist at low energy, our result show the similar competition between these two interactions.

- expectation value $\hat{n}_d$

Hg isotopes are known as nuclei which are located in the transitional region between spherical and deformed shapes. The concept of shape coexistence in this isotopic chain has studied by emphasis on different observables. Here we try to use the expectation value $\hat{n}_d$ which are determined in different states as a new signature for this phenomena. The expectation value of $\hat{n}_d$ is defined as

$$\langle \hat{n}_d \rangle = \frac{\langle [N]\langle N \rangle (\tau) \nu_\Delta LM | \hat{n}_d | [N]\langle N \rangle (\tau) \nu_\Delta LM \rangle}{N}$$

We have determined this quantity in the first states of ground $\langle \hat{n}_d \rangle_g$, $\beta$ $\langle \hat{n}_d \rangle_\beta$ and $\gamma$ $\langle \hat{n}_d \rangle_\gamma$ bands which results are listed in Figure3. Our results suggest an obvious changes in these quantities for two $^{184-186}$Hg isotopes similar what have predicted by Jiao et al [11] which have studied the energy surfaces of Hg and Pt isotopic chains and reported unusual behavior for these isotopes. Also these changes are so remarkable in the $\langle \hat{n}_d \rangle_\beta$ and $\langle \hat{n}_d \rangle_\gamma$ values where the effect of intruder states are increased. The effect of shape coexistence on the excited $0^+$ states are reported on different literatures [10-21] where the related observables such as transition probabilities are studied. The expectation values of $\hat{n}_d$ on these states may be regards as new signature to predict shape coexistence but we need more consideration on the values of these quantities for deformed nuclei to get an exact summary.

- E0 and E2 Transition rates

Stable even-even nuclei in Hg isotopes provide an excellent opportunity for studying the behavior of total low-lying E0 and E2 strengths in the $SU(3) \leftrightarrow \overline{SU(3)}$ transitional region. Computation of electromagnetic transition is a sign of good test for nuclear model wave functions. With using eigenstates which were introduced in Tables (1-4) and Eq.(8-10), the values of different transitions probabilities are determined which are presented in Figures 1a-1f for considered isotopes. Since the experimental data are not available for *E0* transitions in this isotopic chain, we have used these transition to get charge radii which are presented in Table9. The results of present analysis for different quadrupole transition ratios interpret a satisfactory agreement in comparison with experimental counterparts [43-48], too. In all tables of the present paper, the uncertainties of experimental data which are smaller than the size of symbols are not represented.

As have described in Refs.[27-40] about the energy ratio of these nuclei, our results in Figures 1 verify this meaning, a one-parametric Hamiltonian explains very well $R_{4/2}$ ratio on the prolate, SU(3), side of phase transition, i.e. for negative $\chi$ values. At the phase transition and on oblate side, deviations in $R_{4/2}$ ratio are observed. In particular, $^{180-190}_{80}Hg$ isotopes have a slightly smaller $R_{4/2}$ ratio than can be achieved with this simple Hamiltonian. Such structures would have $R_{4/2}$ around or below 2. Instead a slight increase in $R_{4/2}$ suggests an increase in deformation which indicates a deviation from $U(5) \leftrightarrow O(6)$ line towards $SU(3)$. The origin of increased deformation should be related to the quenching of pairing correlations at oblate $Z=80$ and $N=120$ subshells [32-33].

4. Conclusion

In summary, we have studied the energy spectra, monopole and quadrupole transition probabilities, charge radii and quadrupole moments of $^{180-190}$Hg isotopic chain. For this aim, we have determined the SO(6) representation of eigenstates and quadrupole term of transitional Hamiltonian. The results are in the good agreement with the experimental counterparts when the control parameter of Hamiltonian show combination of spherical shapes together deformed ones. Also, the deviation from SO(6) limit is increased when we observe the reduction of quadrupole deformation in these nuclei. The results obtained reinforce this new interpretation of coexistence of shapes or quantum phase transition between prolate and oblate shapes for these nuclei. Also our results for the expectation values of $\hat{n}_d$ operator may realize as new signature of shape coexistence.

**Acknowledgement**

This work is published as a part of research project supported by the University of Tabriz Research Affairs Office.

# Tables

Table1. The SO(6) representation of eigenstates for systems with total boson number $N (= n_s + n_d) = 9$. We have found more than 126 states which the majority of them are not in the experimental spectra. For this aim, we have showed the states which their experimental counterparts are available for considered nuclei.

| $n_d$ | $\sigma$ | $\tau$ | $l$ | Representation |
|---|---|---|---|---|
| 9 | 9 | 7 | 12 | $\sqrt{17/218}[(d^\dagger \times d^\dagger)^4 \times (d^\dagger \times d^\dagger)^4 \times (d^\dagger \times d^\dagger)^4]_m^{12}$ |
| 9 | 9 | 5 | 10 | $\sqrt{11/186}[(d^\dagger \times d^\dagger)^4 \times (d^\dagger \times d^\dagger)^4 \times (d^\dagger \times d^\dagger)^2]_m^{10}$ |
| 9 | 8 | 8 | 12 | $\sqrt{27/185}[(d^\dagger \times d^\dagger)^4 \times (d^\dagger \times d^\dagger)^4 \times (d^\dagger \times d^\dagger)^4]_m^{12}$ |
| 9 | 8 | 6 | 10 | $\sqrt{12/145}[(d^\dagger \times d^\dagger)^4 \times (d^\dagger \times d^\dagger)^4 \times (d^\dagger \times d^\dagger)^2]_m^{10}$ |
| 9 | 6 | 6 | 8 | $\sqrt{16/153}[(d^\dagger \times d^\dagger)^4 \times (d^\dagger \times d^\dagger)^4]_m^{8}$ |
| 8 | 6 | 4 | 8 | $\sqrt{24/127}[(d^\dagger \times d^\dagger)^4 \times (d^\dagger \times d^\dagger)^4]_m^{8}$ |
| 9 | 6 | 6 | 6 | $\sqrt{11/86}[(d^\dagger \times d^\dagger)^4 \times (d^\dagger \times d^\dagger)^4]_m^{6}$ |
| 9 | 4 | 4 | 6 | $\sqrt{10/51}[(d^\dagger \times d^\dagger)^4 \times (d^\dagger \times d^\dagger)^4]_m^{6}$ |
| 8 | 4 | 4 | 6 | $\sqrt{8/45}[(d^\dagger \times d^\dagger)^4 \times (d^\dagger \times d^\dagger)^4]_m^{6}$ |
| 8 | 6 | 4 | 4 | $\sqrt{5/12}(d^\dagger \times d^\dagger)_m^{4}$ |
| 8 | 5 | 4 | 4 | $\sqrt{3/8}(d^\dagger \times d^\dagger)_m^{4}$ |
| 8 | 4 | 2 | 4 | $\sqrt{2/5}(d^\dagger \times d^\dagger)_m^{4}$ |
| 8 | 6 | 6 | 2 | $\sqrt{5/7}(d^\dagger \times d^\dagger)_m^{2}$ |
| 8 | 5 | 4 | 2 | $\sqrt{2/7}(d^\dagger \times d^\dagger)_m^{2}$ |
| 8 | 4 | 2 | 2 | $\sqrt{1/6}(d^\dagger \times d^\dagger)_m^{2}$ |
| 8 | 2 | 2 | 2 | $\sqrt{1/4}(d^\dagger \times d^\dagger)_m^{2}$ |
| 8 | 4 | 4 | 0 | $\sqrt{1/9}(s^\dagger \times d^\dagger)_0^{0}$ |
| 8 | 2 | 4 | 0 | $\sqrt{1/7}(s^\dagger \times d^\dagger)_0^{0}$ |
| 8 | 3 | 2 | 0 | $\sqrt{1/6}(s^\dagger \times d^\dagger)_0^{0}$ |
| 8 | 2 | 2 | 0 | $\sqrt{1/3}(s^\dagger \times d^\dagger)_0^{0}$ |

Table2. The SO(6) representation of eigenstates for systems with total boson number $N (= n_s + n_d) = 10$. We have found more than 170 states which the majority of them are not in the experimental spectra. For this aim, we have showed the states which their experimental counterparts are available for considered nuclei.

| $n_d$ | $\sigma$ | $\tau$ | $l$ | Representation |
|---|---|---|---|---|
| 10 | 10 | 10 | 12 | $\sqrt{24/213}[(d^\dagger \times d^\dagger)^4 \times (d^\dagger \times d^\dagger)^4 \times (d^\dagger \times d^\dagger)^4]_m^{12}$ |
| 10 | 10 | 10 | 10 | $\sqrt{20/99}[(d^\dagger \times d^\dagger)^4 \times (d^\dagger \times d^\dagger)^4 \times (d^\dagger \times d^\dagger)^2]_m^{10}$ |
| 9 | 8 | 8 | 12 | $\sqrt{15/144}[(d^\dagger \times d^\dagger)^4 \times (d^\dagger \times d^\dagger)^4 \times (d^\dagger \times d^\dagger)^4]_m^{12}$ |
| 9 | 8 | 8 | 10 | $\sqrt{16/93}[(d^\dagger \times d^\dagger)^4 \times (d^\dagger \times d^\dagger)^4 \times (d^\dagger \times d^\dagger)^2]_m^{10}$ |
| 9 | 8 | 8 | 8 | $\sqrt{16/87}[(d^\dagger \times d^\dagger)^4 \times (d^\dagger \times d^\dagger)^4]_m^8$ |
| 9 | 6 | 6 | 8 | $\sqrt{13/75}[(d^\dagger \times d^\dagger)^4 \times (d^\dagger \times d^\dagger)^4]_m^8$ |
| 9 | 8 | 8 | 6 | $\sqrt{11/86}[(d^\dagger \times d^\dagger)^4 \times (d^\dagger \times d^\dagger)^4]_m^6$ |
| 9 | 7 | 6 | 6 | $\sqrt{12/35}[(d^\dagger \times d^\dagger)^4 \times (d^\dagger \times d^\dagger)^4]_m^6$ |
| 9 | 7 | 3 | 6 | $\sqrt{9/32}[(d^\dagger \times d^\dagger)^4 \times (s^\dagger \times d^\dagger)^4]_m^6$ |
| 9 | 8 | 6 | 4 | $\sqrt{6/17}(d^\dagger \times d^\dagger)_m^4$ |
| 9 | 7 | 4 | 4 | $\sqrt{5/11}(d^\dagger \times d^\dagger)_m^4$ |
| 9 | 6 | 4 | 4 | $\sqrt{1/5}(d^\dagger \times d^\dagger)_m^4$ |
| 9 | 5 | 6 | 2 | $\sqrt{4/7}(d^\dagger \times d^\dagger)_m^2$ |
| 9 | 4 | 5 | 2 | $\sqrt{1/6}(d^\dagger \times d^\dagger)_m^2$ |
| 9 | 4 | 4 | 2 | $\sqrt{1/5}(s^\dagger \times d^\dagger)_m^2$ |
| 9 | 4 | 3 | 2 | $\sqrt{1/4}(s^\dagger \times d^\dagger)_m^2$ |
| 9 | 5 | 4 | 0 | $\sqrt{1/10}(d^\dagger \times d^\dagger)_0^0$ |
| 9 | 4 | 2 | 0 | $\sqrt{4/9}(s^\dagger \times d^\dagger)_0^0$ |
| 9 | 4 | 0 | 0 | $\sqrt{2/7}(s^\dagger \times d^\dagger)_0^0$ |
| 9 | 3 | 0 | 0 | $\sqrt{2/9}(s^\dagger \times d^\dagger)_0^0$ |

Table3. The SO(6) representation of eigenstates for systems with total boson number $N (= n_s + n_d) = 11$. We have found more than 240 states which the majority of them are not in the experimental spectra. For this aim, we have showed the states which their experimental counterparts are available for considered nuclei.

| $n_d$ | $\sigma$ | $\tau$ | $l$ | Representation |
|---|---|---|---|---|
| 11 | 11 | 10 | 12 | $\sqrt{52/321}[(d^\dagger \times d^\dagger)^4 \times (d^\dagger \times d^\dagger)^4 \times (d^\dagger \times d^\dagger)^4]_m^{12}$ |
| 11 | 10 | 10 | 12 | $\sqrt{39/245}[(d^\dagger \times d^\dagger)^4 \times (d^\dagger \times d^\dagger)^4 \times (d^\dagger \times d^\dagger)^2]_m^{12}$ |
| 10 | 10 | 10 | 10 | $\sqrt{28/192}[(d^\dagger \times d^\dagger)^4 \times (d^\dagger \times d^\dagger)^4 \times (d^\dagger \times d^\dagger)^4]_m^{10}$ |
| 10 | 9 | 8 | 10 | $\sqrt{39/152}[(d^\dagger \times d^\dagger)^4 \times (d^\dagger \times d^\dagger)^4 \times (d^\dagger \times d^\dagger)^2]_m^{10}$ |
| 10 | 8 | 8 | 8 | $\sqrt{22/87}[(d^\dagger \times d^\dagger)^4 \times (d^\dagger \times d^\dagger)^4]_m^{8}$ |
| 10 | 8 | 6 | 8 | $\sqrt{24/75}[(d^\dagger \times d^\dagger)^4 \times (d^\dagger \times d^\dagger)^4]_m^{8}$ |
| 10 | 8 | 8 | 6 | $\sqrt{18/93}[(d^\dagger \times d^\dagger)^4 \times (d^\dagger \times d^\dagger)^4]_m^{6}$ |
| 10 | 6 | 6 | 6 | $\sqrt{8/45}[(d^\dagger \times d^\dagger)^4 \times (d^\dagger \times d^\dagger)^4]_m^{6}$ |
| 10 | 5 | 5 | 6 | $\sqrt{7/25}[(d^\dagger \times d^\dagger)^4 \times (s^\dagger \times d^\dagger)^4]_m^{6}$ |
| 10 | 6 | 6 | 4 | $\sqrt{11/26}(d^\dagger \times d^\dagger)_m^{4}$ |
| 10 | 5 | 4 | 4 | $\sqrt{9/32}(d^\dagger \times d^\dagger)_m^{4}$ |
| 10 | 4 | 4 | 4 | $\sqrt{7/24}(d^\dagger \times d^\dagger)_m^{4}$ |
| 10 | 4 | 6 | 2 | $\sqrt{9/16}(d^\dagger \times d^\dagger)_m^{2}$ |
| 10 | 4 | 5 | 2 | $\sqrt{3/17}(d^\dagger \times d^\dagger)_m^{2}$ |
| 10 | 3 | 4 | 2 | $\sqrt{5/11}(s^\dagger \times d^\dagger)_m^{2}$ |
| 10 | 3 | 3 | 2 | $\sqrt{6/13}(s^\dagger \times d^\dagger)_m^{2}$ |
| 10 | 4 | 2 | 0 | $\sqrt{2/7}(d^\dagger \times d^\dagger)_0^{0}$ |
| 10 | 3 | 0 | 0 | $\sqrt{3/5}(s^\dagger \times d^\dagger)_0^{0}$ |
| 10 | 2 | 2 | 0 | $\sqrt{4/9}(s^\dagger \times d^\dagger)_0^{0}$ |
| 10 | 1 | 0 | 0 | $\sqrt{1/3}(s^\dagger \times d^\dagger)_0^{0}$ |

Table4. The SO(6) representation of eigenstates for systems with total boson number $N (= n_s + n_d) = 12$. We have found more than 350 states which the majority of them are not in the experimental spectra. For this aim, we have showed the states which their experimental counterparts are available for considered nuclei.

| $n_d$ | $\sigma$ | $\tau$ | $l$ | Representation |
|---|---|---|---|---|
| 12 | 10 | 10 | 12 | $\sqrt{78/345}[(d^\dagger \times d^\dagger)^4 \times (d^\dagger \times d^\dagger)^4 \times (d^\dagger \times d^\dagger)^4]^{12}_m$ |
| 12 | 10 | 10 | 12 | $\sqrt{63/144}[(d^\dagger \times d^\dagger)^4 \times (d^\dagger \times d^\dagger)^4 \times (d^\dagger \times d^\dagger)^2]^{12}_m$ |
| 11 | 8 | 10 | 10 | $\sqrt{55/126}[(d^\dagger \times d^\dagger)^4 \times (d^\dagger \times d^\dagger)^4 \times (d^\dagger \times d^\dagger)^4]^{10}_m$ |
| 11 | 8 | 9 | 10 | $\sqrt{16/93}[(d^\dagger \times d^\dagger)^4 \times (d^\dagger \times d^\dagger)^4 \times (d^\dagger \times d^\dagger)^2]^{10}_m$ |
| 11 | 7 | 6 | 8 | $\sqrt{18/85}[(d^\dagger \times d^\dagger)^4 \times (d^\dagger \times d^\dagger)^4]^8_m$ |
| 11 | 6 | 6 | 8 | $\sqrt{13/75}[(d^\dagger \times d^\dagger)^4 \times (d^\dagger \times d^\dagger)^4]^8_m$ |
| 11 | 7 | 7 | 6 | $\sqrt{14/45}[(d^\dagger \times d^\dagger)^4 \times (d^\dagger \times d^\dagger)^4]^6_m$ |
| 11 | 6 | 6 | 6 | $\sqrt{9/22}[(d^\dagger \times d^\dagger)^4 \times (d^\dagger \times d^\dagger)^4]^6_m$ |
| 11 | 6 | 5 | 6 | $\sqrt{12/45}[(d^\dagger \times d^\dagger)^4 \times (s^\dagger \times d^\dagger)^4]^6_m$ |
| 11 | 7 | 4 | 4 | $\sqrt{12/35}(d^\dagger \times d^\dagger)^4_m$ |
| 11 | 6 | 3 | 4 | $\sqrt{9/26}(d^\dagger \times d^\dagger)^4_m$ |
| 11 | 5 | 2 | 4 | $\sqrt{7/15}(d^\dagger \times d^\dagger)^4_m$ |
| 11 | 4 | 3 | 2 | $\sqrt{8/17}(d^\dagger \times d^\dagger)^2_m$ |
| 11 | 3 | 3 | 2 | $\sqrt{11/28}(d^\dagger \times d^\dagger)^2_m$ |
| 11 | 4 | 4 | 2 | $\sqrt{3/11}(s^\dagger \times d^\dagger)^2_m$ |
| 11 | 2 | 2 | 2 | $\sqrt{3/8}(s^\dagger \times d^\dagger)^2_m$ |
| 11 | 6 | 4 | 0 | $\sqrt{4/9}(d^\dagger \times d^\dagger)^0_0$ |
| 11 | 4 | 2 | 0 | $\sqrt{5/7}(s^\dagger \times d^\dagger)^0_0$ |
| 11 | 3 | 0 | 0 | $\sqrt{2/5}(s^\dagger \times d^\dagger)^0_0$ |
| 11 | 2 | 0 | 0 | $\sqrt{1/4}(s^\dagger \times d^\dagger)^0_0$ |

Table 5. Analytical expression of quadrupole operator, $Q^\chi$, for systems with N=9.

| L | $\varepsilon$ |
|---|---|
| 0 | $24510 + \frac{1204}{11}\chi^2 + \frac{3118}{17}\chi^4 + \frac{820}{114}\chi^6 + \frac{22054}{6}\chi^8 + \frac{71036}{55}\chi^{10} + \frac{30025}{87}\chi^{12}$ |
| 2 | $1204 + \frac{431}{105}\chi^2 + \frac{706}{149}\chi^4 + \frac{3971}{354}\chi^6 + \frac{231}{2105}\chi^8 + \frac{588}{7851}\chi^{10}$ |
| 4 | $658 + \frac{2685}{18}\chi^2 + \frac{1460}{28}\chi^4 + \frac{1059}{49}\chi^6 + \frac{605}{1250}\chi^8 + \frac{75}{1980}\chi^{10}$ |
| 6 | $127 + \frac{341}{14}\chi^2 + \frac{2980}{289}\chi^4 + \frac{1455}{686}\chi^6$ |
| 8 | $35 + \frac{104}{15}\chi^2 + \frac{26}{75}\chi^4 + \frac{52}{225}\chi^6$ |
| 10 | $94 + \frac{179}{13}\chi^2 + \frac{71}{150}\chi^4 + \frac{53}{440}\chi^6$ |
| 12 | $201 + \frac{98}{5}\chi^2 + \frac{158}{45}\chi^4$ |

Table 6. Analytical expression of quadrupole operator, $Q^\chi$, for systems with N=10.

| L | $\varepsilon$ |
|---|---|
| 0 | $3447 + \frac{875}{18}\chi^2 + \frac{1785}{42}\chi^4 + \frac{1255}{279}\chi^6 + \frac{45772}{25}\chi^8 + \frac{61005}{42}\chi^{10} + \frac{14550}{51}\chi^{12}$ |
| 2 | $2300 + \frac{21}{88}\chi^2 + \frac{766}{231}\chi^4 + \frac{327}{75}\chi^6 + \frac{1440}{243}\chi^8 + \frac{755}{5572}\chi^{10}$ |
| 4 | $24 + \frac{745}{12}\chi^2 + \frac{1220}{33}\chi^4 + \frac{966}{85}\chi^6 + \frac{1026}{645}\chi^8 + \frac{142}{2781}\chi^{10}$ |
| 6 | $52 + \frac{196}{21}\chi^2 + \frac{984}{85}\chi^4 + \frac{2054}{1101}\chi^6$ |
| 8 | $71 + \frac{85}{39}\chi^2 + \frac{145}{111}\chi^4 + \frac{282}{135}\chi^6$ |
| 10 | $46 + \frac{92}{25}\chi^2 + \frac{85}{76}\chi^4 + \frac{110}{325}\chi^6$ |
| 12 | $247 + \frac{84}{27}\chi^2 + \frac{212}{81}\chi^4$ |

Table 7. Analytical expression of quadrupole operator, $Q^\chi$, for systems with N=11.

| L | $\varepsilon$ |
|---|---|
| 0 | $5400 + \frac{342}{25}\chi^2 + \frac{955}{39}\chi^4 + \frac{1755}{123}\chi^6 + \frac{12550}{112}\chi^8 + \frac{22120}{552}\chi^{10} + \frac{47680}{1461}\chi^{12}$ |
| 2 | $4788 + \frac{75}{42}\chi^2 + \frac{651}{122}\chi^4 + \frac{855}{1441}\chi^6 + \frac{3220}{2439}\chi^8 + \frac{4881}{7800}\chi^{10}$ |
| 4 | $166 + \frac{561}{24}\chi^2 + \frac{869}{45}\chi^4 + \frac{1022}{483}\chi^6 + \frac{1775}{1128}\chi^8 + \frac{2445}{3441}\chi^{10}$ |
| 6 | $133 + \frac{96}{45}\chi^2 + \frac{807}{150}\chi^4 + \frac{1773}{2552}\chi^6$ |
| 8 | $202 + \frac{144}{75}\chi^2 + \frac{855}{372}\chi^4 + \frac{1065}{836}\chi^6$ |
| 10 | $102 + \frac{65}{42}\chi^2 + \frac{35}{122}\chi^4 + \frac{455}{766}\chi^6$ |
| 12 | $366 + \frac{144}{45}\chi^2 + \frac{608}{123}\chi^4$ |

Table 8. Analytical expression of quadrupole operator, $Q^\chi$, for systems with N=12.

| L | $\varepsilon$ |
|---|---|
| 0 | $3220 + \frac{155}{18}\chi^2 + \frac{1125}{88}\chi^4 + \frac{144}{1005}\chi^6 + \frac{18650}{5472}\chi^8 + \frac{2806}{11450}\chi^{10} + \frac{112470}{2601}\chi^{12}$ |
| 2 | $6440 + \frac{844}{155}\chi^2 + \frac{1665}{753}\chi^4 + \frac{2980}{2502}\chi^6 + \frac{17500}{4602}\chi^8 + \frac{34006}{12033}\chi^{10}$ |
| 4 | $1055 + \frac{166}{75}\chi^2 + \frac{355}{124}\chi^4 + \frac{805}{612}\chi^6 + \frac{2305}{891}\chi^8 + \frac{5788}{1350}\chi^{10}$ |
| 6 | $452 + \frac{114}{27}\chi^2 + \frac{1335}{108}\chi^4 + \frac{840}{1404}\chi^6$ |
| 8 | $96 + \frac{582}{265}\chi^2 + \frac{1455}{822}\chi^4 + \frac{471}{155}\chi^6$ |
| 10 | $85 + \frac{114}{35}\chi^2 + \frac{372}{85}\chi^4 + \frac{1445}{1203}\chi^6$ |
| 12 | $14 + \frac{275}{18}\chi^2 + \frac{925}{62}\chi^4$ |

Table 9. Charge radii (in fm$^2$) and quadrupole moment (in W.u.) of isotopic chain in the 2$^+$ state of ground bound. Experimental counterparts are taken from Ref.[87] and [91] for charge radii and quadrupole moments, respectively.

| Nucleus | $\delta\langle r^2 \rangle_{Exp.}$ | $\delta\langle r^2 \rangle_{Theo.}$ | $Q_2^{Exp.}$ | $Q_2^{Theo.}$ |
|---|---|---|---|---|
| $^{180}$Hg | - 0.527* | - 0.562 | - 0.192 | - 0.207 |
| $^{182}$Hg | - 0.693 | - 0.712 | - 0.225 | - 0.241 |
| $^{184}$Hg | - 0.550 | - 0.569 | - 0.293 | - 0.305 |
| $^{186}$Hg | - 0.477 | - 0.481 | - 0.155 | - 0.173 |
| $^{188}$Hg | - 0.404 | - 0.419 | - 0.098 | - 0.107 |
| $^{190}$Hg | - 0.326 | - 0.334 | - 0.082 | - 0.094 |

- This data is taken from Ref.[13].

Figure Caption

Figure1a. The energy levels (in keV) and transition probabilities (in W.u.) of $^{180}$Hg. The parameters of transitional Hamiltonian are $\eta = 0.69$, $E_0 = 455$ and $C = 2.55$ keV and parameters of quadrupole transition operator are e = 2.541 and $\chi = -0.117$.

Figure1b. The energy levels (in keV) and transition probabilities (in W.u.) of $^{182}$Hg. The parameters of transitional Hamiltonian are $\eta = 0.84$, $E_0 = 597$ and $C = 2.37$ keV and parameters of quadrupole transition operator are e = 2.011 and $\chi = -0.097$.

Figure1c. The energy levels (in keV) and transition probabilities (in W.u.) of $^{184}$Hg. The parameters of transitional Hamiltonian are $\eta = 0.91$, $E_0 = 618$ and $C = 2.15$ keV and parameters of quadrupole transition operator are e = 2.214 and $\chi = -0.103$.

Figure1d. The energy levels (in keV) and transition probabilities (in W.u.) of $^{186}$Hg. The parameters of transitional Hamiltonian are $\eta = 0.64$, $E_0 = 551$ and $C = 2.31$ keV and parameters of quadrupole transition operator are e = 2.001 and $\chi = -0.088$.

Figure1e. The energy levels (in keV) and transition probabilities (in W.u.) of $^{188}$Hg. The parameters of transitional Hamiltonian are $\eta = 0.78$, $E_0 = 533$ and $C = 2.27$ keV and parameters of quadrupole transition operator are e = 1.994 and $\chi = -0.079$.

Figure1f. The energy levels (in keV) and transition probabilities (in W.u.) of $^{190}$Hg. The parameters of transitional Hamiltonian are $\eta = 0.83$, $E_0 = 566$ and $C = 2.14$ keV.

Figure2. Variation of control parameter ($\eta$) versus quadrupole deformation ($\beta_2$) for considered nuclei.

Figure3. Variation of the expectation values of $\hat{n}_d$ operator which are determined in the first states of ground $\langle \hat{n}_d \rangle_g$, $\beta$ $\langle \hat{n}_d \rangle_\beta$ and $\gamma$ $\langle \hat{n}_d \rangle_\gamma$ bands for considered nuclei. .

Figure1a.

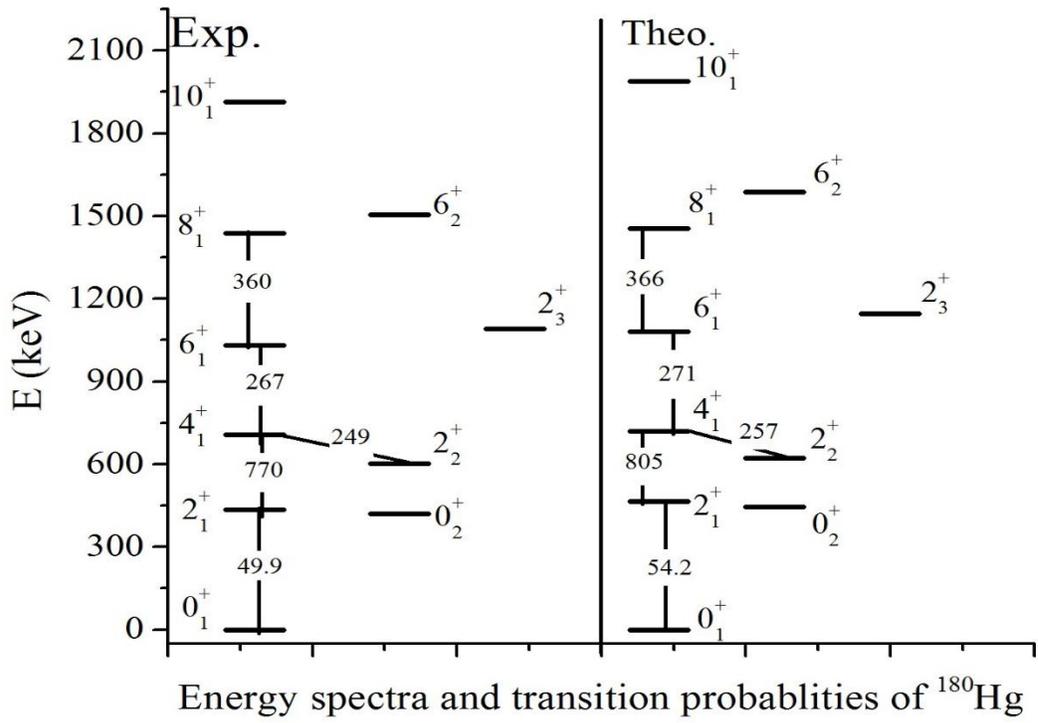

Figure1b.

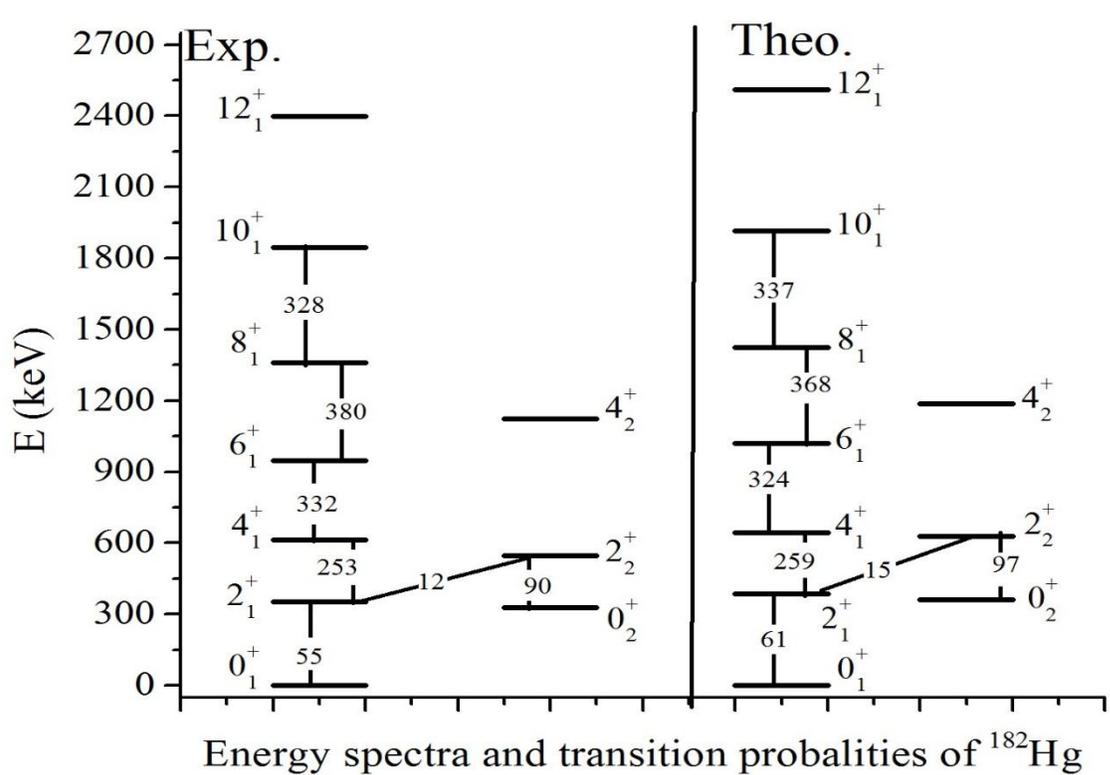

Figure1c.

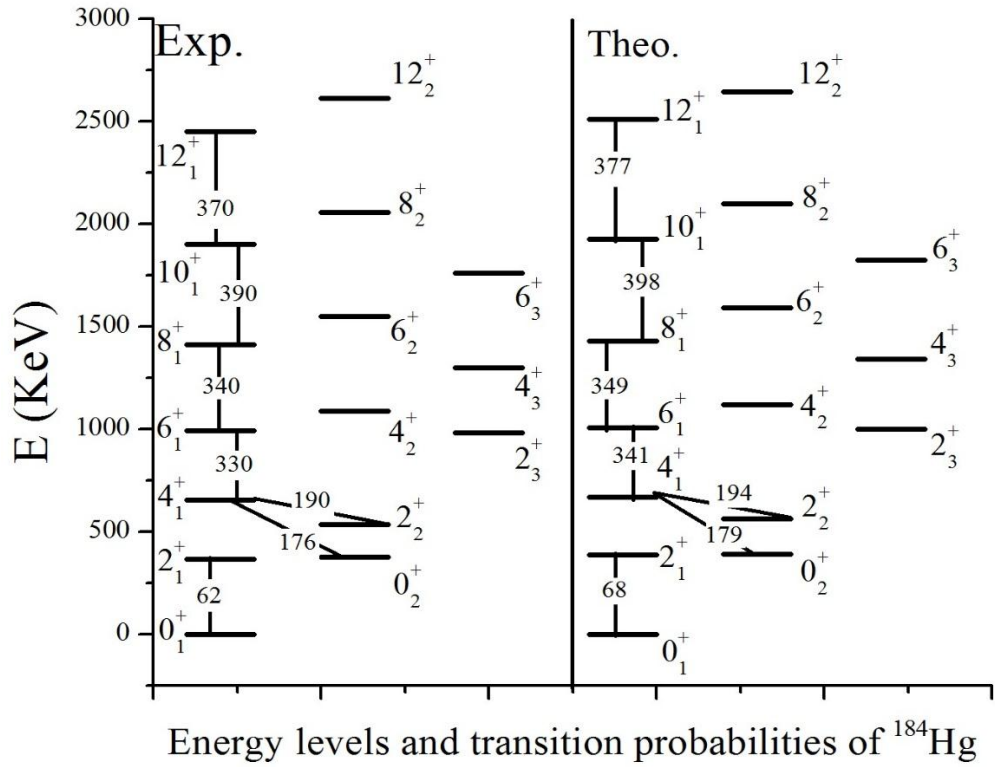

Energy levels and transition probabilities of $^{184}$Hg

Figure1d.

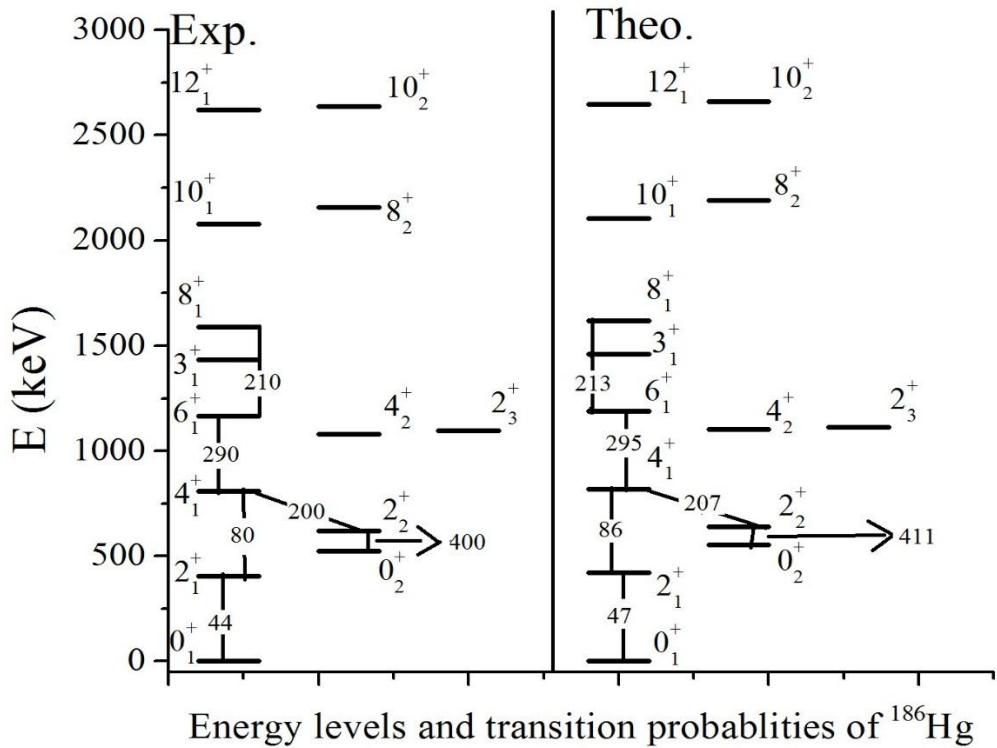

Energy levels and transition probablities of $^{186}$Hg

Figure1e.

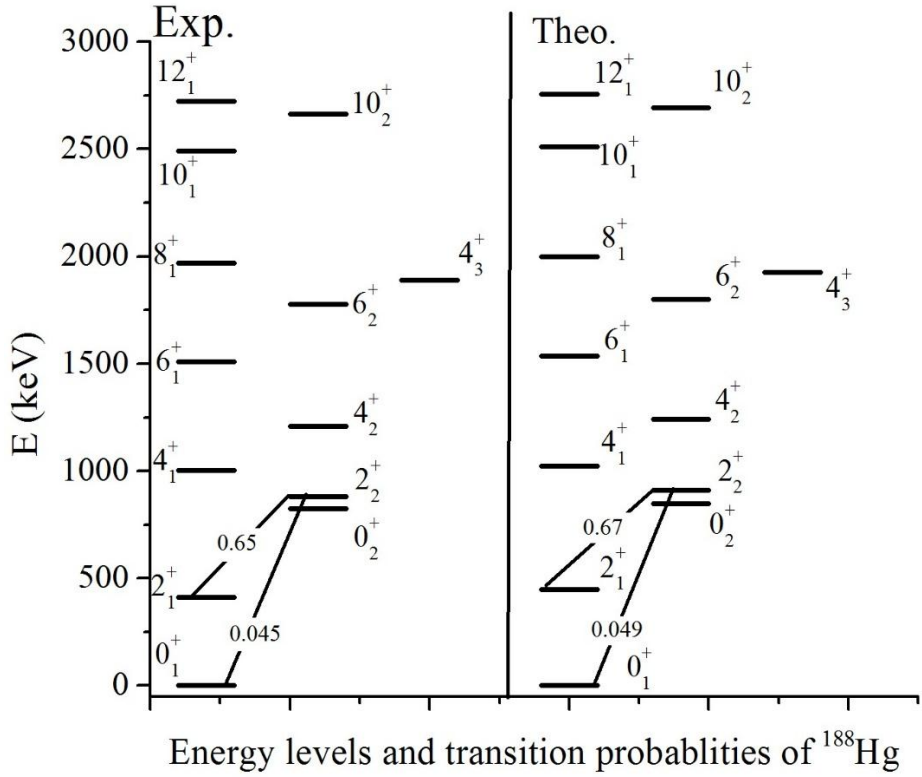

Energy levels and transition probablities of $^{188}$Hg

Figure1f.

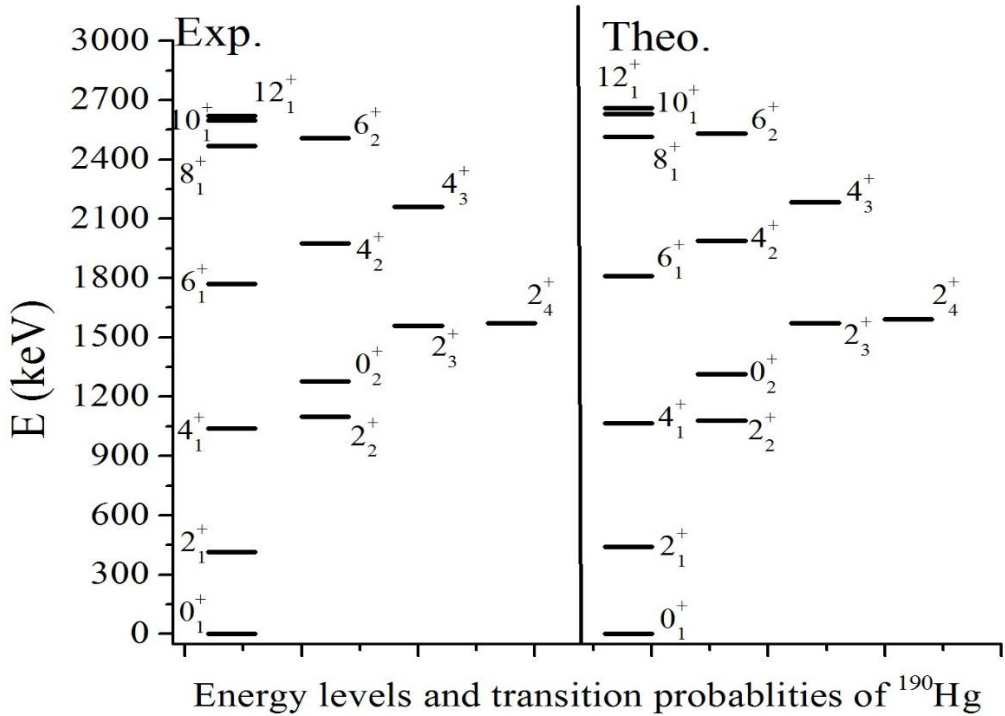

Energy levels and transition probablities of $^{190}$Hg

Figure2.

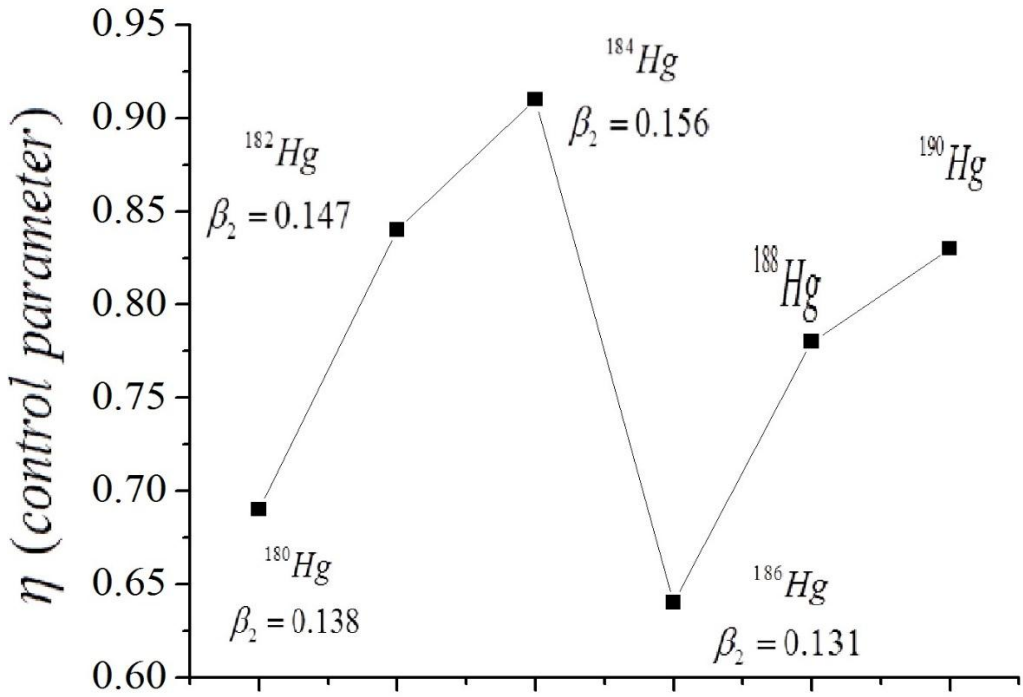

Figure3.

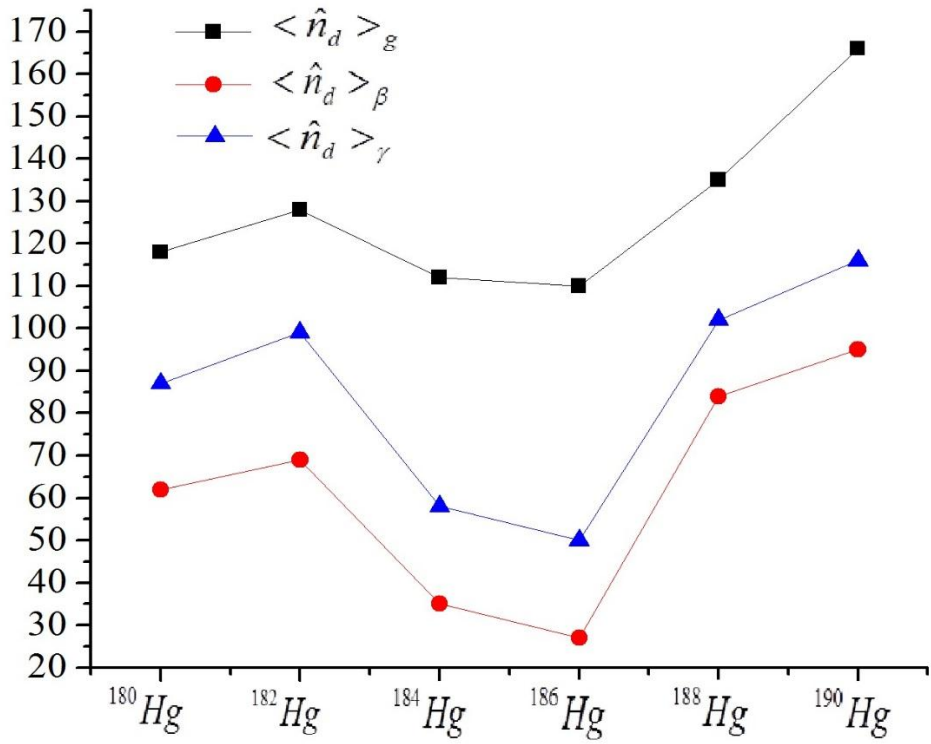